\documentclass[apj]{emulateapj}

\def\lea{\mathrel{<\kern-1.0em\lower0.9ex\hbox{$\sim$}}}
\def\gea{\mathrel{>\kern-1.0em\lower0.9ex\hbox{$\sim$}}}

\slugcomment{}

\shorttitle{The formation of massive cluster galaxies}
\shortauthors{Mancone, Gonzalez, Brodwin, Stanford, Eisenhardt, Stern, Jones}

\begin{document}


\title{The Formation of Massive Cluster Galaxies}

\author{Conor L. Mancone\altaffilmark{1},
Anthony H. Gonzalez\altaffilmark{1},
Mark Brodwin\altaffilmark{2},
Spencer A. Stanford\altaffilmark{3} \altaffilmark{4},
Peter R. M. Eisenhardt\altaffilmark{5},
Daniel Stern\altaffilmark{5},
Christine Jones\altaffilmark{2}}
\altaffiltext{1}{Department of Astronomy, University of Florida, Gainesville, FL 32611}
\altaffiltext{2}{Harvard-Smithsonian Center for Astrophysics, 60 Garden Street, Cambridge, MA 02138}
\altaffiltext{3}{Physics Department, University of California, Davis, CA 95616}
\altaffiltext{4}{Institute of Geophysics and Planetary Physics, Lawrence Livermore National Laboratory, Livermore, CA 94550}
\altaffiltext{5}{Jet Propulsion Laboratory, California Institute of Technology, 4800 Oak Grove Drive, Pasadena, CA 91109}
\email{cmancone@astro.ufl.edu, anthony@astro.ufl.edu}



\begin{abstract}
We present composite 3.6 and 4.5$\mu$m luminosity functions for cluster galaxies measured from the Spitzer Deep, Wide-Field Survey (SDWFS) for $0.3<z<2$. We compare the evolution of $m^*$ for these luminosity functions to models for passively evolving stellar populations to constrain the primary epoch of star formation in massive cluster galaxies. At low redshifts ($z \lesssim 1.3$) our results agree well with models with no mass assembly and passively evolving stellar populations with a luminosity-weighted mean formation redshift $z_f=2.4$ assuming a Kroupa initial mass function (IMF). We conduct a thorough investigation of systematic biases that might influence our results, and estimate systematic uncertainites of $\Delta z_f=^{+0.16}_{-0.18}$ (model normalization), $\Delta z_f=^{+0.40}_{-0.05}$ ($\alpha$), and $\Delta z_f=^{+0.30}_{-0.45}$ (choice of stellar population model). For a Salpeter type IMF, the typical formation epoch is thus strongly constrained to be $z\sim2-3$. Higher formation redshifts can only be made consistent with the data if one permits an evolving IMF that is bottom-light at high redshift, as suggested by \citet{vanDokkum08}. At high redshift ($z \gtrsim 1.3$) we also witness a statistically significant ($>5\sigma$) disagreement between the measured luminosity function and the continuation of the passive evolution model from lower redshifts. After considering potential systematic biases that might influence our highest redshift data points, we interpret the observed deviation as potential evidence for ongoing mass assembly at this epoch.
\end{abstract}



\keywords{galaxies: clusters, galaxies: evolution, galaxies: formation, galaxies: luminosity function}

\defcitealias{CB07}{CB07}
\defcitealias{M08}{M08}
\defcitealias{BC03}{BC03}
\defcitealias{M05}{M05}
\defcitealias{D09}{D09}
\defcitealias{lin04}{L04}

\section{Introduction}

Two fundamental aspects of galaxy formation are the star formation history and the mass assembly history.  In principle the star formation history is easier to measure because the effects of star formation remain visible even long after it has ceased.  In contrast, the signatures of mass assembly are typically either short-lived or difficult to extract once the galaxy returns to virial equilibrium.  This means that while the star formation history can be measured for individual galaxies, the mass assembly history is best measured from a population of galaxies, and these galaxies must be observed while they are still assembling.

Studies of clusters often treat the galaxies as if they formed their stars in one short burst at high redshift, and so parameterize their star formation histories by the luminosity-weighted or mass-weighted mean age of their stellar populations.  While this is certainly not true in general (see for instance \citealt{trager08}) it is a useful simplification and works surprisingly well, especially for massive cluster galaxies.

A number of different techniques have been used to measure the mean age of the stellar populations in galaxy clusters.  The most straight-forward technique is to use spectral indices or model comparisons to galaxy spectra to measure ages directly.  \citet{thomas05} study early type galaxies in dense environments using such a method and find that the majority of the star formation in massive galaxies occurred between $z \sim3 - 5$, with vigorous star formation episodes from $z \sim2 - 5$.  Age estimates for cluster galaxies can also come from the cluster red sequence in various ways.  \citet{kurk09} uses the color of the red sequence in a single $z = 1.6$ assembling cluster to estimate a formation epoch of $z_f \sim 3$.  \citet{tran07} examine the small scatter of absorption line galaxies about the red sequence of a z=0.83 galaxy cluster, and conclude that all star formation ceased by $z \sim 1.2$ and that most of the members formed by $z \sim 2$.  This is consistent with their own estimate of the member ages from the color of the red sequence.

Meanwhile, the fundamental plane can be used to measure the ages of stellar populations by observing the evolution in M/L ratios as a function of redshift.  Performing such an analysis with three galaxy clusters at $z \sim 0.5$, \citet{vanDokkum07} estimate a luminosity-weighted mean formation redshift for their galaxies of $2.01^{+0.22}_{-0.17}$.  For surveys that have many clusters distributed over a range of redshifts, it is possible to examine the color or luminosity evolution of cluster galaxies over time to constrain the primary star formation epoch.  \citet{eisenhardt} explore a sample of clusters with photometric redshifts between $0 < z < 2.4$ and find color evolution that is consistent with $z_f \geq 3$ (with larger formation redshifts preferred for $z > 1$ clusters).  \citet[hereafter M08]{M08} find a cluster sample with photometric redshifts between $0.1 < z < 1.3$ by searching for red sequence galaxies, and conclude that the luminosity evolution of their sample is consistent with passively evolving galaxies formed at $z > 1.5$.

It is important to note that luminosity-weighted measures of stellar population ages will generally return later formation redshifts (and younger ages) than mass-weighted measurements, since young stellar populations contribute more luminosity per unit mass than old stellar populations.  We note that of the papers quoted above, \citet{thomas05} is the only one that provides a mass-weighted age measurement, and so it is not surprising that it points to the earliest formation redshifts.  It is hard to say whether the observed variation between the luminosity-weighted age estimates represents genuine differences in the stellar populations, differences in the methods, or signs of systematic errors.  One recognized uncertainty that factors into many of these studies is the slope of the IMF (in particular near $1M_\odot$) at high redshift.  Since the rate of luminosity evolution is strongly dependent upon the slope of the IMF \citep{conroy}, any of the above studies that estimate formation epochs based on the evolution of galaxy luminosities (e.g. \citealt{vanDokkum07}, \citetalias{M08}) are strongly dependent upon the slope of the IMF.  In general flatter IMFs at earlier epochs would push these studies to earlier formation redshifts.

There is still much work needed on our picture of mass assembly in the most massive cluster galaxies.  Past work on the luminosity evolution in cluster galaxies has typically searched for deviations of the galaxies relative to models for passively evolving galaxies.  Finding none, studies have concluded that there is little or no mass assembly in massive cluster galaxies, out to the farthest redshifts studied, $z \lesssim 1.3$ (see \citealt{strazzullo}, \citealt{depropris}, \citetalias{M08}).  However it is not clear just how much mass assembly is ruled out by such studies.  Agreement with passive evolution models represents, at best, an upper limit on the amount of mass assembly allowed, but this limit has not been explicitly calculated.  Another method of constraining mass assembly is to look at the build-up of the red sequence.  The red sequence has the advantage of being relatively easy to identify, even at high redshift.  Indeed, \citet{kodama07} find evidence that the massive end of the red sequence begins to build up as early as $2 \lesssim z \lesssim 3$, and \citet{zirm08} find a red sequence around a $z = 2.16$ protocluster.  This suggests that the bright end of the cluster red sequence is established very early, a fact verified by well defined red sequences found in two assembling clusters, one at $z = 1.6$ \citep{kurk09} and another at $z = 1.62$ \citep{papovich10, tanaka10}.

In this paper we extend the work of previous researchers who have studied the evolution of the luminosity function of massive cluster galaxies (\citealt{strazzullo}, \citealt{depropris}, \citealt{muzzin07}, \citetalias{M08}).  We make use of the Spitzer Deep, Wide-Field Survey (SDWFS, \citealt{ashby}), an 8.5 deg$^2$ survey with a limiting magnitude of 18.8 ($5\sigma$) at $4.5\mu$m (on the Vega-mag system).  This survey has sufficient multiwavelength data to calculate accurate photometric redshifts and reliably identify clusters and likely members.  Because of the size of our survey area we have a larger cluster sample than previous work in the literature, and the depth of our survey allows us to find clusters out to high redshifts ($0 < z < 2.4$).

This paper is structured as follows: Section 2 describes the data we are working with, Section 3 presents our luminosity functions, and in Section 4 we compare them to models for passively evolving stellar populations.  In Section 5 we discuss the evidence for mass assembly in our results, and Section 6 contains our summary and conclusions.  Throughout this work we assume a WMAP5 cosmology, $\Omega_0 = 0.279$, $\Omega_\Lambda = 0.721$, and $H_0 = 70.1$ km s$^{-1}$Mpc$^{-1}$ \citep{wmap5}.  All magnitudes are on the Vega magnitude system.

\section{Data}

\subsection{Galaxy Catalog}

Our galaxy catalog for this work comes from SDWFS, which is a reimaging of the IRAC Shallow Cluster Survey (ISCS, \citealt{eisenhardt04}).  The ISCS had 90 seconds of integration time across $8.5$deg$^2$ in the Bo\"{o}tes field, and SDWFS added 3 more 90 second exposures at every pointing.  \citet{ashby} describes the data reduction for SDWFS.  The source catalog reaches depths of $19.77$ and $18.83$ mags in $3.6$ and $4.5\mu$m and contains 670,446 detected sources at $3.6\mu$m and 528,232 sources at $4.5\mu$m ($5\sigma$, 4" aperture corrected to total).  We use 4" aperture magnitudes corrected to total to derive photometric redshift probability distribution functions from combined SDWFS ($3.6$, $4.5$, $5.8$, and $8.0\mu$m) imaging and $B_wRI$ data from the NOAO Deep Wide-Field Survey (NDWFS).  We work with a subsample of the SDWFS catalog consisting of sources which are brighter than $18.8$ mags in $4.5\mu$m and have optical data.  Our subsample consists of 454,418 sources for which we assign redshift probability distributions using the methodology of \citet{brodwin2006}.  Comparing to 15,052 galaxies with spectroscopic redshifts, \citet{brodwin2006} find that their photometric redshifts are good to $\Delta z = 0.06(1+z)$ for $95\%$ of their galaxies at $0 < z < 1.2$.  We find a similar quality using the deeper SDWFS data.  From simulations we expect our photometric redshifts to remain robust out to $z \sim 3$.  Photometric redshift errors remain as low as $\sigma_z/(1+z) = 0.06$ out to $z = 1.2$, and increases to a maximum error of $\sigma_z/(1+z) \sim 0.2$ at $z = 1.7$, dropping back down to $\sigma_z/(1+z) \sim 0.15$ at $z = 2.0$.  This has been verified with spectroscopic redshifts out to $z \lesssim 1.5$.

\subsection{Cluster Catalog}

We utilize a sample of 335 clusters from the ISCS \citep{stanford,elston,brodwin2006,eisenhardt,stern10}.  The identification of the cluster sample is described in \citet{eisenhardt}, and we summarize the major features here.  This cluster list was generated by using the redshift probability distributions from \citet{brodwin2006} to perform a wavelet analysis and identify clusters in the ISCS.  With this method probability maps are generated for fixed redshift slices, and ``signal" is added to the probability maps at the location of the galaxies in proportion to the integrated probability of each galaxy being found in the redshift slice.  Clusters are then identified by searching for statistically significant peaks in these probability maps.  Of the 335 clusters in this survey 25\% are confirmed, including 15 at $z>1$ \citep{stanford, eisenhardt}.  We restrict our analysis to $0.3 < z < 2.0$ limiting our sample to 296 clusters of which 20\% are confirmed.  Confirmation of low redshift clusters comes primarily from the AGES survey \citep{kochanek04}.  Since redshift probability distributions were the basis for this cluster search, the cluster catalog should be relatively robust against assumptions of galaxy type, the existence of a red sequence, or any other requirements typically adopted by other cluster finding algorithms.  Finally, we emphasize that this work relies on improved photmetric redshifts for our galaxies calculated from NDWFS+SDWFS data, though our cluster catalog was generated using photometric redshifts from the shallower NDWFS+ISCS data.

\section{Observed Luminosity Function}

\subsection{General Procedure}

While using photometric redshifts allows us to construct a large galaxy catalog out to high redshifts, it is not without its disadvantages.  Without spectroscopic redshifts we can only find cluster members in a statistical fashion.  Instead of measuring the luminosity function of cluster galaxies directly, we must measure the luminosity function of galaxies near each cluster and correct for the presence of field galaxies.  To accomplish this we first find photo-z cluster members and field galaxies for each cluster.  Galaxies within 1.5 Mpc (physical distance) and near the cluster redshift are considered photo-z cluster members, and galaxies with $4$ Mpc $ < d < 8$ Mpc and near the cluster redshift are considered field galaxies.  We then measure the field luminosity function by fitting a \citet{schechter} luminosity function to the field galaxies near each cluster.  We parameterize the luminosity function of the photo-z cluster members as the sum of another Schechter (the cluster luminosity function) and the field luminosity function.  For the actual fitting process we divide the clusters into redshift bins and calculate the $k$-correction and distance modulus correction needed to move a passively evolving galaxy from the cluster redshift (using spectroscopic redshifts for the clusters when available) to the bin redshift.  We apply these corrections to the member and field galaxies for each cluster.  We then apply the above procedure to the combined photo-z cluster members and field galaxies for all the clusters in each redshift bin.

\subsection{Fitting Details\label{fitting_details}}

While the above outline describes our general fitting procedure, there are a number of important details that go into the fitting process.  The first consideration is how to determine whether or not a particular galaxy is at the cluster redshift.  For this we use the full redshift probability distribution functions for each galaxy.  We calculate the probability that each galaxy would be found within $\pm 0.06(1+z)$ of the cluster redshift.  Galaxies that fall below a certain probability threshold are removed from the fitting process.  We find that requiring a total probability $>30\%$ provides a balance between removing field galaxies without cutting too many cluster galaxies.  However, we also redo our fitting process using cuts at 20, 40, and 50\% to investigate the impact this has on our results (see Sections \ref{lf_errors} and \ref{fitting_errors}).  Our choice of cut probability has only a minor impact on our results.  Using a fixed cut at all redshifts will remove a larger fraction of galaxies at high redshift, because their photometric redshift probability distributions are broader.  To account for this we fix the cut probability at $z=0$ and allow it to decrease as a function of cluster redshift to account for the broadening of the photometric redshifts to high redshift.  The cut decreases such that a galaxy that is at the cluster redshift and makes the cut at $z=0$ will not be excluded if it is at the redshift of a higher-z cluster.  Also, this probability cut disproportionately removes faint galaxies because they have the largest photo-z errors at a given redshift.  We correct for this effect by performing a simple Monte Carlo simulation using the redshift probability distributions for the galaxies to estimate the fraction of galaxies lost as a function of magnitude, due to this probability cut.  We then weight the galaxies appropriately in our fits to account for this incompleteness.

We must account for any potential overlap between clusters.  We search for any clusters that fall within 10 Mpc and $\Delta z = 0.06(1+z)$ of a given cluster, and remove any photo-z cluster members or field galaxies that are within 2 Mpc of a contaminating cluster.  We also calculate the relative size of the field and member regions for each cluster, accounting for any missing areas due to contaminating clusters, bad pixels, bright stars, or the edge of the survey.

The clusters are divided into redshift bins with a width of $\Delta z=0.15$ starting at $z=0.3$.  Instead of considering the center of each bin to be the bin location, we use the median redshift of the clusters in each bin as the bin location.  We combine high redshift bins to increase signal to noise, so our two highest redshift bins go from $z = 1.35 - 1.63$ and $z = 1.63 - 2.00$.  We remove approximately 12\% of the clusters in our sample because a neighboring cluster is located within 1.5 Mpc and $\Delta z = \pm 0.06(1+z)$.  We exclude another 3\% because the density of galaxies near the cluster is no greater than in the field.  Some of these are caused by large-scale structures in the field region, and the rest are likely to be false positives.  Of the original 296 clusters we exclude a total of 46 for these various reasons, leaving 250 clusters which go into our fitting procedure.

We parameterize the field luminosity function (fLF) as a Schechter function but our redshift cut should select a population of normally distributed galaxies with $\sigma = 0.06(1+z)$ centered at the cluster redshift.  This means that the fLF should be similar to a Schechter distribution except smoothed with a Gaussian kernel on a scale determined by our redshift selection criteria.  For any individual cluster this smoothing length should be close to the total $k$-correction and distance modulus correction needed to move a galaxy from $\pm 0.06(1+z)$ to the cluster redshift.  So for the fit to the fLF in a redshift bin we use a Schechter function smoothed by a Gaussian kernel.  For each cluster we calculate the $k$-correction and distance modulus correction needed to go from the cluster redshift to $\pm 0.06(1+z)$.  The average of these values for the clusters in a redshift bin becomes the scale length of the Gaussian used to smooth the fLF.  Typical values for this smoothing length vary from $\sim 0.2 - 0.5$ mags, and are larger at lower redshifts.  Given this smoothing length, we fit a smoothed Schechter function to the fLF in each redshift bin.  We perform a simple $\chi^2$ minimization to the binned fLF, using a binsize of $0.3$ mags.  In addition the contribution of each field galaxy to the luminosity function is weighted by the ratio of the cluster area to field area for the cluster it originated from.  We parameterize the Schechter function as:

\begin{equation} N(m) = 0.4\ln(10)\Phi^*10^{-0.4(m-m^*)(\alpha+1)}\exp(-10^{-0.4(m-m^*)}) \end{equation}

While fitting both the cluster and field LF, we correct for photometric incompleteness from the SDWFS survey by weighting galaxies according to the mean incompleteness measured as a function of magnitude \citep{ashby}.  For reference the completeness remains above $75\%$ ($70\%$) for 3.6 (4.5)$\mu$m for all our galaxies.  For measuring both the field and cluster luminosity function, fitting is limited to galaxies brighter than the apparent magnitude limit of each redshift bin.  The apparent magnitude limit of each bin is taken to be 18.7 (18.8) mags in 3.6 (4.5)$\mu$m, plus the minimum $k$-correction plus distance modulus correction applied to any cluster in that redshift bin.  This correction is a function of redshift and varies between about $-1.0$ mag in the low redshift bins to $\sim-0.2$ mags at the high redshift end.

We fit for the cluster luminosity function by summing a Schechter function and the fLF and fitting this to the luminosity function of the photo-z cluster members in each redshift bin.  We use maximum likelihood fitting as an alternative to a binned fit, following the general procedure of \citet{marshall}.  Our data are only deep enough to reliably constrain $\alpha$ in the lowest redshift bin, for which we find $\alpha = -0.6 \pm 0.2$.  Previous work in the literature has commonly found \citep[hereafter L04]{lin04} or assumed \citepalias{M08} $\alpha \sim -0.8$ for clusters.  This is consistent with the value from our lowest redshift bin, so we also assume and fix $\alpha = -0.8$ in our fits for all redshift bins.  In Section \ref{alpha_errors} we explore the sensitivity of our results to different values of $\alpha$.  We use a downhill simplex algorithm \citep{nr} to maximize the likelihood as a function of $m^*$ and $\Phi^*$.  The relationship between the likelihood, $m^*$, and $\Phi^*$ is smooth with only one maximum, so our results are independent of the starting guess.

Finally, we estimate errors in $m^*$ for each redshift bin using bootstrap resampling.  We generate realizations of each redshift bin by randomly selecting N clusters from the bin with replacement, where N is equal to the number of clusters in the bin.  This number of realizations for each bin is such that every redshift bin will have at least $N\log(N)^2$ realizations, which has been shown to be a sufficient number for accurately estimating errors \citep{babu}.  We fit each realization in the same fashion as before.  Error estimates come from the distribution of fitted $m^*$ values to the realizations.  We find the $m^*$ values that contain $68\%$ of the fitted distribution, and use these as our errors.

\subsection{Statistical Background Subtraction\label{stat_subtract}}

We also measure the cluster LF using a more typical statistical background subtraction to verify that there are no large systematic errors in our results.  We use the same cluster and field regions as before and create binned cluster and field LFs for each cluster, using a binsize of 0.2 mags.  The fLF is then subtracted from the cluster luminosity function.  We bin the clusters in redshift space, this time with a binsize of $\Delta z = 0.2$, and apply $k$-corrections and distance modulus corrections to move the LFs from the cluster redshift to the center of each bin.  The field subtracted LFs are then added together in a redshift bin, and a Schechter function is fit using Levenberg-Marquardt least-squares minimization.  Errors are derived with bootstrap resampling.

\subsection{Results}

\begin{figure*}
\epsscale{0.9}
\plotone{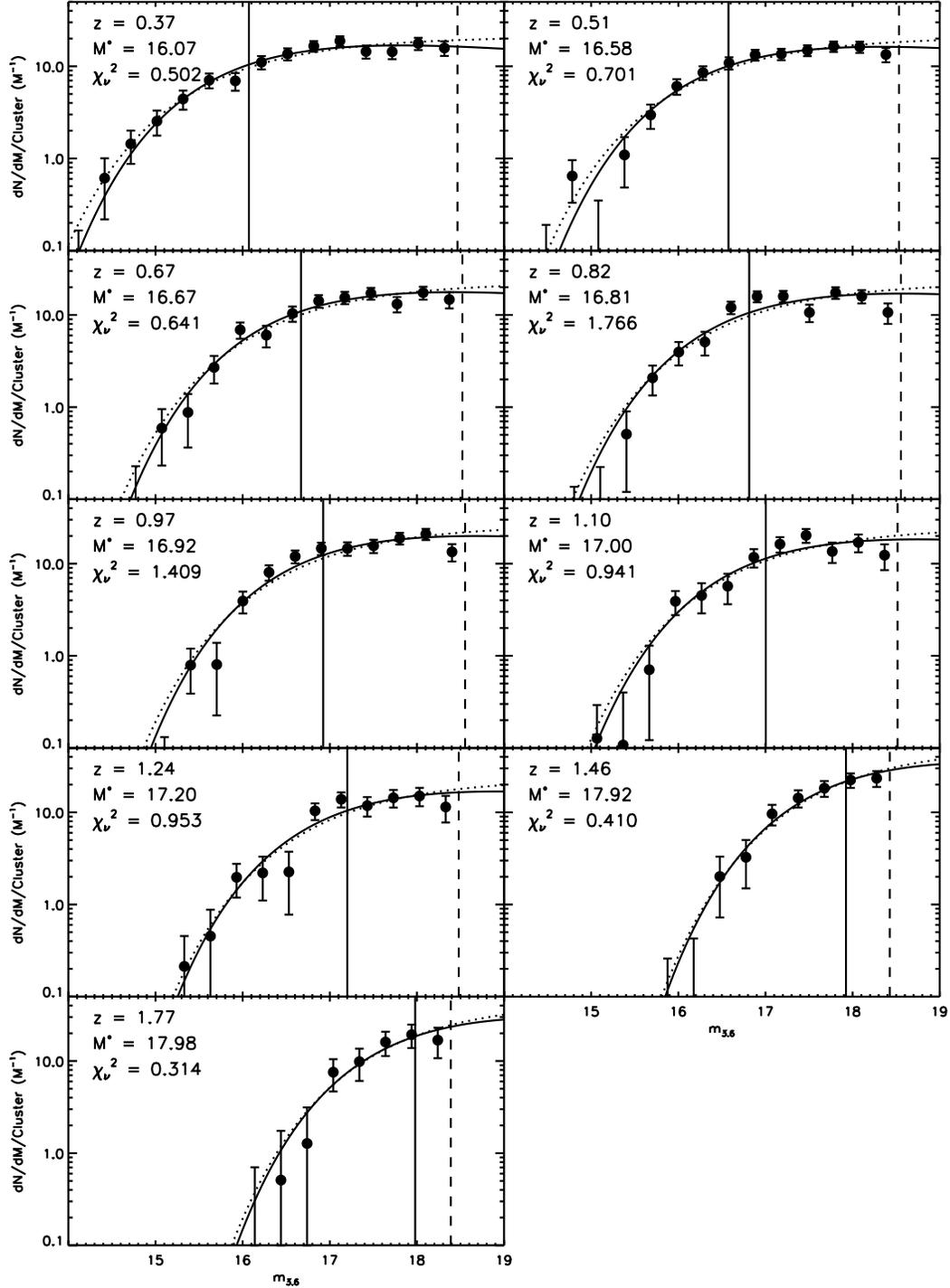}
\caption{Schechter fits to the $3.6\mu$m cluster luminosity function in each redshift bin.  The solid circles are binned differences between the luminosity function for all galaxies within 1.5 Mpc of the cluster center at the cluster redshift and the area-weighted luminosity function of the field in units of galaxies per magnitude per cluster.  The solid curve shows the best fitting Schechter function and the solid vertical line shows the fitted value of $m_{3.6}^*$ assuming $\alpha = -0.8$ in all redshift bins.  The dotted curve shows the fit with $\alpha = -1.0$.  The dashed vertical line denotes the apparent magnitude limit in each bin.}\label{fig:fitssumch1}
\end{figure*}

\begin{figure*}
\epsscale{0.9}
\plotone{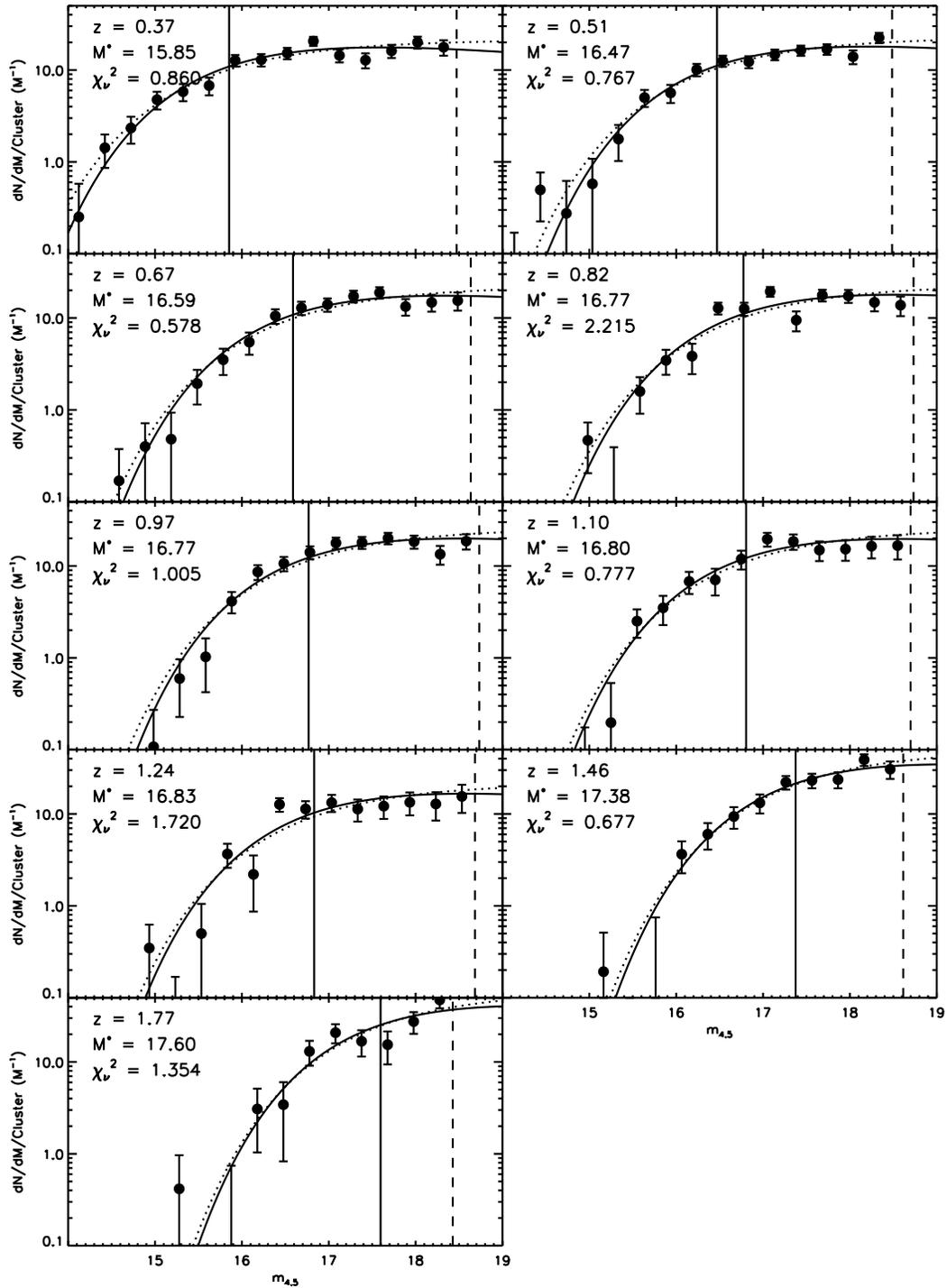}
\caption{The same as Figure \ref{fig:fitssumch1}, but for $4.5\mu$m.}\label{fig:fitssumch2}
\end{figure*}

The fits to the cluster luminosity functions for each redshift bin are shown in Figures \ref{fig:fitssumch1} (3.6$\mu$m) and \ref{fig:fitssumch2} (4.5$\mu$m).  For clarity we plot in Figures \ref{fig:fitssumch1} and \ref{fig:fitssumch2} the (binned) difference between the luminosity function of galaxies near the cluster and the field luminosity function in units of galaxies per magnitude.  This is represented by solid circles.  The solid curve denotes the Schechter function with $\alpha = -0.8$ fitted to the cluster luminosity function, and the solid vertical line shows the fitted value of $m^*$.  The dotted line shows the fitted Schechter function with $\alpha = -1.0$, and the dashed vertical line is the apparent magnitude limit of the redshift bin.  Figures \ref{fig:fitssumch1} and \ref{fig:fitssumch2} show that our fitted values of $m^*$ are $\sim$2.5 mags brighter than the apparent magnitude limit of the survey at low redshift.  At high redshift the 4.5$\mu$m data are $\sim$1 mag brighter than the apparent magnitude limit, while the $3.6\mu$m data are only 0.5 mags brighter.  Table \ref{tbl:results} lists the best fitting $m^*$ values and errors for $\alpha = -0.6, -0.8, -1.0,$ and $-1.12$ as a function of redshift.  Table \ref{tbl:results_stats} lists the results calculated with our statistical background subtraction.

\begin{deluxetable}{ccccc}
\tablecaption{Schechter Fit Paramaters\\Photo-z Selection\label{tbl:results}}
\tablewidth{0pt}
\tablehead{
  \colhead{$\alpha$} & \colhead{z} & \colhead{\# Clusters} & \colhead{$M^*_{3.6}$} & \colhead{$M^*_{4.5}$}
}
\startdata
-0.60 & 0.37 & 35 & 16.30$^{+0.12}_{-0.10}$ & 16.09$^{+0.11}_{-0.10}$ \\
-0.60 & 0.51 & 44 & 16.78$^{+0.10}_{-0.07}$ & 16.67$^{+0.08}_{-0.07}$ \\
-0.60 & 0.67 & 29 & 16.86$^{+0.10}_{-0.09}$ & 16.82$^{+0.10}_{-0.10}$ \\
-0.60 & 0.82 & 35 & 17.00$^{+0.08}_{-0.06}$ & 16.97$^{+0.06}_{-0.07}$ \\
-0.60 & 0.97 & 30 & 17.11$^{+0.07}_{-0.06}$ & 16.96$^{+0.05}_{-0.05}$ \\
-0.60 & 1.10 & 21 & 17.18$^{+0.13}_{-0.09}$ & 16.99$^{+0.07}_{-0.10}$ \\
-0.60 & 1.24 & 25 & 17.37$^{+0.14}_{-0.12}$ & 16.98$^{+0.09}_{-0.10}$ \\
-0.60 & 1.46 & 22 & 18.04$^{+0.10}_{-0.06}$ & 17.53$^{+0.08}_{-0.07}$ \\
-0.60 & 1.77 &  9 & 18.11$^{+0.38}_{-0.20}$ & 17.73$^{+0.20}_{-0.09}$ \\
      &      &    &                         &                         \\
-0.80 & 0.37 & 35 & 16.07$^{+0.13}_{-0.12}$ & 15.85$^{+0.10}_{-0.12}$ \\
-0.80 & 0.51 & 44 & 16.58$^{+0.08}_{-0.07}$ & 16.48$^{+0.08}_{-0.09}$ \\
-0.80 & 0.67 & 29 & 16.67$^{+0.09}_{-0.12}$ & 16.59$^{+0.10}_{-0.09}$ \\
-0.80 & 0.82 & 35 & 16.81$^{+0.09}_{-0.05}$ & 16.77$^{+0.07}_{-0.10}$ \\
-0.80 & 0.97 & 30 & 16.92$^{+0.06}_{-0.06}$ & 16.77$^{+0.08}_{-0.05}$ \\
-0.80 & 1.10 & 21 & 17.00$^{+0.08}_{-0.12}$ & 16.79$^{+0.10}_{-0.10}$ \\
-0.80 & 1.24 & 25 & 17.20$^{+0.14}_{-0.14}$ & 16.81$^{+0.09}_{-0.07}$ \\
-0.80 & 1.46 & 22 & 17.92$^{+0.09}_{-0.10}$ & 17.38$^{+0.09}_{-0.12}$ \\
-0.80 & 1.77 &  9 & 17.98$^{+0.42}_{-0.23}$ & 17.60$^{+0.16}_{-0.18}$ \\
      &      &    &                         &                         \\
-1.00 & 0.37 & 35 & 15.79$^{+0.10}_{-0.14}$ & 15.54$^{+0.10}_{-0.13}$ \\
-1.00 & 0.51 & 44 & 16.32$^{+0.10}_{-0.10}$ & 16.20$^{+0.09}_{-0.07}$ \\
-1.00 & 0.67 & 29 & 16.44$^{+0.09}_{-0.11}$ & 16.36$^{+0.11}_{-0.12}$ \\
-1.00 & 0.82 & 35 & 16.62$^{+0.10}_{-0.07}$ & 16.55$^{+0.08}_{-0.08}$ \\
-1.00 & 0.97 & 30 & 16.73$^{+0.07}_{-0.05}$ & 16.55$^{+0.07}_{-0.06}$ \\
-1.00 & 1.10 & 21 & 16.81$^{+0.13}_{-0.12}$ & 16.61$^{+0.08}_{-0.12}$ \\
-1.00 & 1.24 & 25 & 17.04$^{+0.11}_{-0.11}$ & 16.63$^{+0.09}_{-0.10}$ \\
-1.00 & 1.46 & 22 & 17.80$^{+0.12}_{-0.10}$ & 17.22$^{+0.12}_{-0.09}$ \\
-1.00 & 1.77 &  9 & 17.84$^{+0.29}_{-0.39}$ & 17.45$^{+0.18}_{-0.12}$ \\
      &      &    &                         &                         \\
-1.12 & 0.37 & 35 & 15.60$^{+0.14}_{-0.12}$ & 15.34$^{+0.12}_{-0.13}$ \\
-1.12 & 0.51 & 44 & 16.16$^{+0.11}_{-0.08}$ & 16.03$^{+0.10}_{-0.07}$ \\
-1.12 & 0.67 & 29 & 16.29$^{+0.11}_{-0.10}$ & 16.21$^{+0.13}_{-0.10}$ \\
-1.12 & 0.82 & 35 & 16.48$^{+0.11}_{-0.09}$ & 16.40$^{+0.08}_{-0.09}$ \\
-1.12 & 0.97 & 30 & 16.60$^{+0.12}_{-0.07}$ & 16.38$^{+0.07}_{-0.07}$ \\
-1.12 & 1.10 & 21 & 16.68$^{+0.14}_{-0.13}$ & 16.44$^{+0.11}_{-0.09}$ \\
-1.12 & 1.24 & 25 & 16.90$^{+0.20}_{-0.10}$ & 16.51$^{+0.11}_{-0.12}$ \\
-1.12 & 1.46 & 22 & 17.72$^{+0.11}_{-0.08}$ & 17.09$^{+0.08}_{-0.09}$ \\
-1.12 & 1.77 &  9 & 17.74$^{+0.45}_{-0.34}$ & 17.36$^{+0.20}_{-0.14}$ \\
\enddata
\end{deluxetable}

\begin{deluxetable}{ccccc}
\tablecaption{Schechter Fit Paramaters\\Statistical Background Subtraction\label{tbl:results_stats}}
\tablewidth{0pt}
\tablehead{
  \colhead{$\alpha$} & \colhead{z} & \colhead{\# Clusters} & \colhead{$M^*_{3.6}$} & \colhead{$M^*_{4.5}$}
}
\startdata
-0.80 & 0.55 & 43 & 16.67 $\pm$ 0.07 & 16.57 $\pm$ 0.11 \\
-0.80 & 0.79 & 35 & 16.91 $\pm$ 0.10 & 16.80 $\pm$ 0.08 \\
-0.80 & 0.98 & 37 & 16.99 $\pm$ 0.09 & 16.79 $\pm$ 0.09 \\
-0.80 & 1.20 & 27 & 17.22 $\pm$ 0.17 & 16.87 $\pm$ 0.19 \\
-0.80 & 1.41 & 17 & 17.67 $\pm$ 0.22 & 17.05 $\pm$ 0.21 \\
-0.80 & 1.60 & 13 & 17.82 $\pm$ 0.23 & 17.30 $\pm$ 0.14 \\
-0.80 & 1.77 &  3 & 16.91 $\pm$ 0.57 & 16.41 $\pm$ 0.68 \\
\enddata
\end{deluxetable}

\subsection{Fitting Errors\label{lf_errors}}

There are multiple potential systematic uncertainties related to the LF fitting procedure.  The first comes from our choice of probability cut for determining which galaxies are at the cluster redshift.  As mentioned in Section \ref{fitting_details} we reject galaxies which have a total probability $<30\%$ of being within $\Delta z = 0.06(1+z)$ of the cluster redshift.  Using a probability cut of $20\%$ moves our results brighter by $\sim 0.05$ mags, and cutting at $40\%$ moves our results fainter by $\sim 0.05$ mags.  A cut of $50\%$ scatters our results randomly by $\sim 0.1$ mags.

Second is the luminosity function completeness correction which we have applied to account for the loss of faint galaxies due to our probability cut.  If we do not include this correction, our fitted $m^*$ values move systematically brighter by $\sim 0.05$ mags.

Third, our choice of $\alpha$ is an important source of fitting uncertainty in this study.  For now we note that adjusting $\alpha$ produces a largely systematic shift in our results.  In general if $\alpha$ becomes steeper by $0.1$ then our fitted $m^*$ values move brighter by $\sim0.1$ mags.  We discuss the impact of our choice of $\alpha$ in more detail in Section \ref{alpha_errors} because comparison to our models for passive galaxy evolution provides the best way to investigate the impact of $\alpha$ on our results.

\section{Models for Passively Evolving Stellar Populations}

\subsection{Model Description}

\begin{figure}
\epsscale{1.0}
\plotone{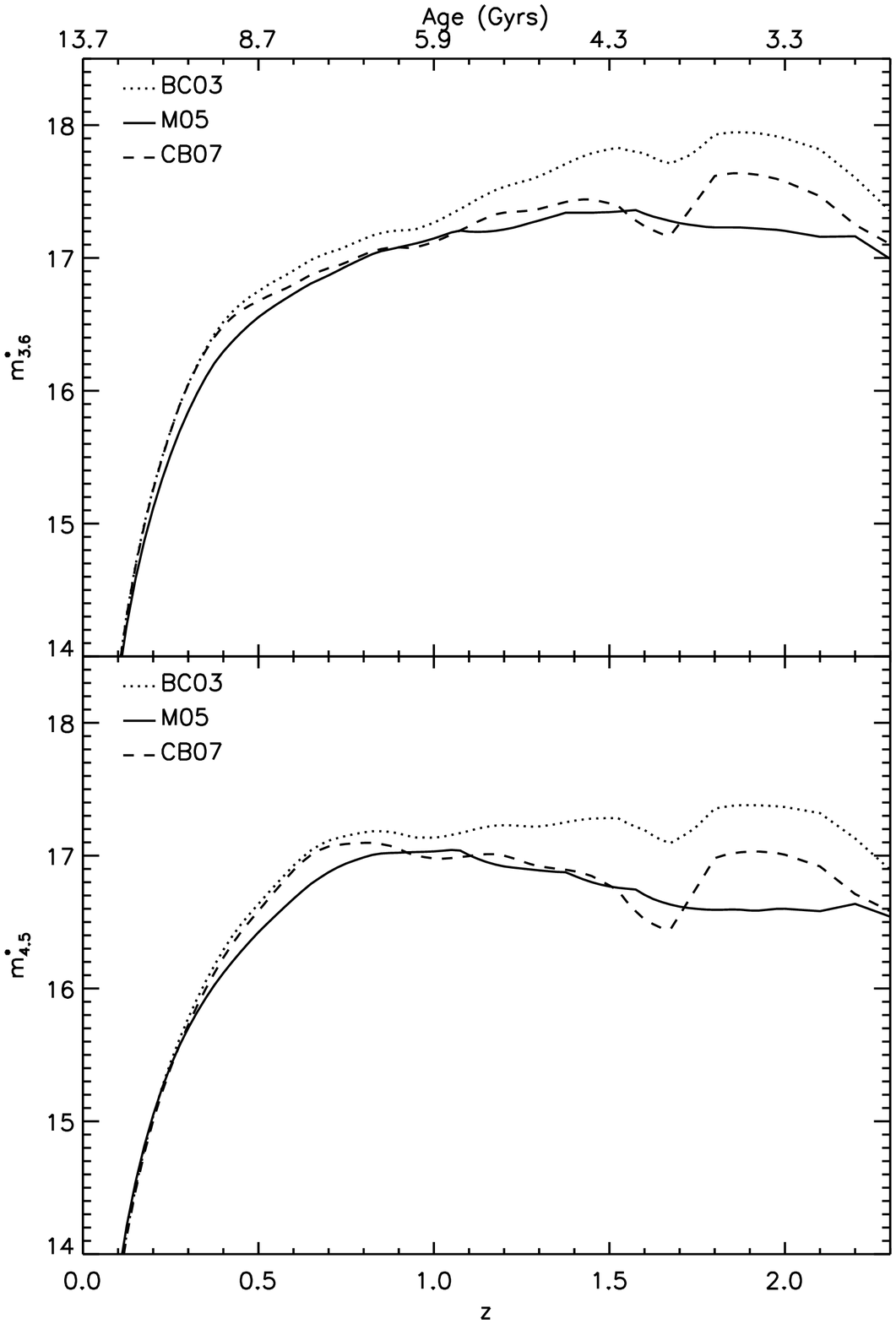}
\caption{A comparison between the models of \protect\citetalias{BC03} (dotted line), \protect\citetalias{CB07} (dashed line), and \protect\citetalias{M05} (solid line).  The star formation epoch is assumed to be $z=2.5$ for all models.  Redshift is on the x-axis.  Top panel is $3.6\mu$m, bottom panel is $4.5\mu$m.  All three model sets are normalized to match the results of \protect\citetalias{lin04} at low redshift.}\label{fig:comparemodels}
\end{figure}

To measure the primary star formation epoch for galaxies in this work we compare our measured $m^*$ values against model predictions for a passively evolving stellar population.  We compare the predictions for three different stellar model sets, \citet[hereafter BC03]{BC03}, an updated version of \citetalias{BC03} with detailed treatment of thermally-pulsing asymptotic giant branch (TP-AGB) stars \citep[hereafter CB07]{CB07}, and the models from \citet[hereafter M05]{M05}.  Figure \ref{fig:comparemodels} compares the predictions of these three model sets at $3.6\mu$m (top), and $4.5\mu$m (bottom).  In this figure all the models are normalized to have the same $M^*$ value at $z = 0.05$ in $3.6\mu$m ($M_{3.6}^*$), which we discuss in more detail below.  \citetalias{CB07} and \citetalias{BC03} use a \citet{chabrier} IMF and \citetalias{M05} uses a Kroupa IMF.  For all three model sets we assume solar metallicity, and in Section \ref{sps_errors} we investigate the results of changing IMF or metallicity.  We characterize the star formation history as an exponentially declining burst of star formation with a characteristic timescale of 0.1 Gyr, effectively a single burst model.  Since our goal is to compare our results against models for passively evolving stellar populations, we concentrate on this one ``bursting" model of star formation and do not attempt to model more complicated star formation histories.  We compare all three model sets in Figure \ref{fig:comparemodels}, which shows the predicted evolution in $3.6\mu$m when star formation turns on at $z=2.5$.  The \citetalias{BC03} and \citetalias{CB07} models agree well at low redshift, but start to diverge at higher redshifts.  The \citetalias{CB07} models are brighter than the \citetalias{BC03} models at higher redshift, due primarily to a more improved treatment of TP-AGB stars in the \citetalias{CB07} models.  \citetalias{M05}, which also includes an updated treatment of TP-AGB stars, is brighter than \citetalias{CB07} at $z \sim 0.6$, and lacks the dip at $z \sim 1.8$ that both \citetalias{CB07} and \citetalias{BC03} have.  For the remainder of this paper we focus upon \citetalias{M05} but also compare our results with \citetalias{BC03} and \citetalias{CB07} (Section \ref{sps_errors}) to determine the systematic errors introduced by our choice of model.

\begin{figure}
\epsscale{0.8}
\plotone{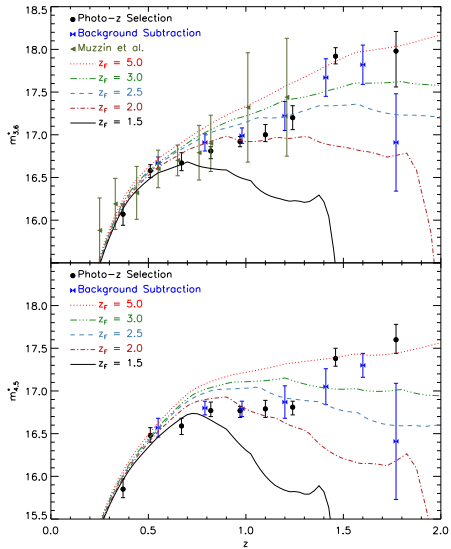}
\caption{Measured evolution of $m_{3.6}^*$ (top) and $m_{4.5}^*$ (bottom) versus redshift compared with \protect\citetalias{M05} model predictions.  We show our results from both our photo-z selection method and statistical background subtraction.  The models are normalized to match the results of \protect\citetalias{lin04} at low redshift.  Also included in the top plot are the results of \protect\citetalias{M08} who measured $m_{3.6}^*$ as a function of redshift.}\label{fig:results}
\end{figure}

With all three model sets we build a range of models which differ only by their peak star formation epoch.  We use peak star formation redshifts that span the range of $z_f=1.0$ to $z_f=9.95$ in redshift increments of $\Delta z = 0.05$.  Previous studies have used measurements of the galaxy luminosity function at low redshift to normalize such models out to high redshift.  However our data are of sufficient quality that we can fit for both the model normalization and formation redshift at the same time, allowing for a more self-consistent result.  For all of our model fits we will let the model normalization be a free parameter.  However in some cases (such as Figure \ref{fig:comparemodels}) we show models without having performed a fit to the data.  For simplicity in these cases, we fix the model normalization using the results from \citetalias{lin04} for which we calculate $M_{3.6}^* = -24.32$ for galaxies at a mean redshift of $z = 0.05$.  This comes from taking the original result found by \citetalias{lin04} in $K$s band ($M_{Ks}^* = -24.02$) and using our models to convert from rest-frame $K$s to observed $3.6\mu$m.

\subsection{Model Comparison to Cluster LF\label{model_comparison}}

\begin{figure}
\epsscale{1.0}
\plotone{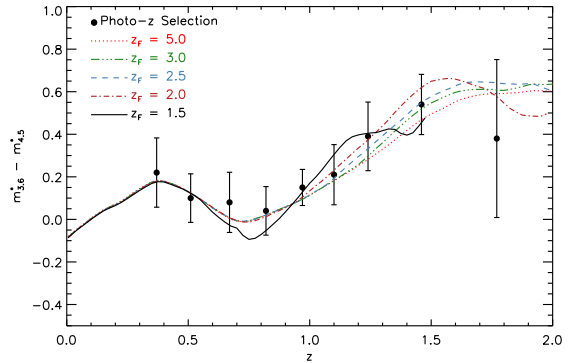}
\caption{Measured evolution of $m_{3.6}^*$ - $m_{4.5}^*$ versus redshift compared with \protect\citetalias{M05} model predictions.}\label{fig:resultscolor}
\end{figure}

The measured $m^*$ values from our photo-z selection method with $\alpha$ fixed at $-0.8$ are plotted against redshift in Figure \ref{fig:results}.  A small subsample of the passive evolution models are overplotted (with the normalization fixed according to \citetalias{lin04}).  Figure \ref{fig:results} includes the results of \citetalias{M08}.  Our error bars are smaller as expected because of our use of photometric redshifts to select galaxies (as opposed to the statistical background subtraction used by \citetalias{M08}), and because of the larger area and greater depth of the SDWFS survey.  \citetalias{M08} used data from the Spitzer First Look Survey \citep{fls}, a $3.8$ deg$^2$ field with 1/6th of our integration time.  In Figure \ref{fig:results} we also include the results of our statistical background subtraction and find excellent agreement between our photo-z selection method and our statistical background subtraction.  For this reason we will concentrate on the results from our photo-z selection method and only return to our statistical background subtraction when investigating the impact of systematic uncertainties (Section \ref{fitting_errors}).  Figure \ref{fig:resultscolor} compares our results with the predicted color evolution.  The color evolution is not useful for constraining the formation epoch because there is little difference between the models.  Instead we note that the color evolution we observe is consistent with the model predictions and hence serves as a useful check of our results.

\begin{figure}
\epsscale{1.0}
\plotone{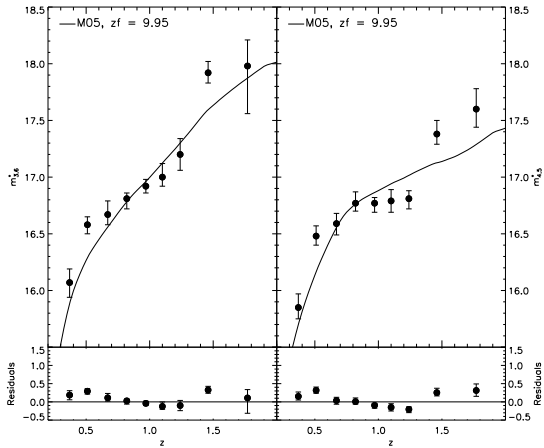}
\caption{Best fitting model using the entire cluster sample out to $z = 1.86$.  Redshift is plotted versus $m_{3.6}^*$ (top left) and $m_{4.5}^*$ (top right).  The solid line is the best fitting model for passively evolving stellar populations with $z_f = 9.95$.  The models are normalized to have $M_{3.6}^* = -24.71$ at z = 0.05, as determined from the fitting process.  The circles correspond to the measured evolution of $m_*$ with $\alpha$ fixed to $-0.8$.  Differences between the data and the models are plotted for $3.6\mu$m (bottom left) and $4.5\mu$m (bottom right).}\label{fig:bestfit}
\end{figure}

To measure the actual formation redshift we perform a simple $\chi^2$ minimization between the data and the models, allowing the model normalization to float.  The best fit model has $z_f = 9.95$, which is the earliest formation redshift we attempted to fit.  $\chi_{\nu}^2$ for this best fit is 4.63, which excludes the best fitting model for passive evolution at $> 5\sigma$.  The reason for this discrepancy is illustrated by comparing our observed evolution with the best fitting model, shown in Figure \ref{fig:bestfit}.  The top left and top right plots show the observed evolution and best fitting model with $z_f = 9.95$ for $3.6$ and $4.5\mu$m, respectively.  The lower left and lower right panels show the residuals of the fit.  The large $\chi^2$ is clearly caused because the $z > 1.4$ data points are faint relative to the rest of the data, and so the models miss both the low redshift ($z < 1.3$) and high redshift ($z > 1.3$) data.  A close inspection of Figure \ref{fig:results} shows that no combination of formation redshift and model normalization can match both the low redshift and high redshift data.

\begin{figure}
\epsscale{1.0}
\plotone{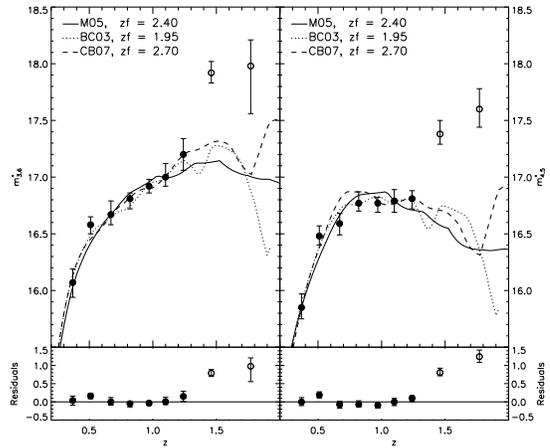}
\caption{Same as Figure \ref{fig:bestfit} except the two highest redshift bins (open circles) have been excluded from the fit.  The new best fit model has $z_f = 2.40$ and $M_{3.6}^* = -24.47$.}\label{fig:best_fit_lowz}
\end{figure}

For this reason we redo the $\chi^2$ fit without the two highest redshift data points.  The resulting fit is plotted in Figure \ref{fig:best_fit_lowz}.  The preferred formation redshift is now $z_f = 2.4^{+0.16}_{-0.18}$.  The quality of the fit is substantially improved ($\chi_{\nu}^2 = 1.5$).  An analysis of $\chi^2$ residuals reveals that both the high redshift $3.6\mu$m data and the $4.5\mu$m data reject the best fitting $z_f = 2.40$ model at very high confidence ($>5\sigma$).  There are two plausible reasons for this discrepancy at high redshifts.  First, there is always the possibility that some systematic error becomes important at high redshifts.  Second is that this is a signature of galaxy assembly, which would cause $M^*$ to grow fainter at higher redshifts as the galaxies break up into their progenitors.  We explore both of these possibilities more in Section \ref{mass_assembly}.  Either way we will exclude the highest redshift data points from all further analysis involving our models for passive stellar evolution.

\subsection{Systematic Uncertainties\label{model_errors}}

There are a number of systematic uncertainties that affect our ability to accurately measure the formation redshift.  We explore these in detail below.

\subsubsection{Model Normalization ($M_{3.6}^*$)\label{norm_errors}}

\begin{deluxetable*}{cccccccccccc}
\tablecaption{Model normalizations from the literature.\label{tbl:normalizations}}
\tablewidth{0pt}
\tablehead{
  \colhead{Reference}
  &\colhead{Band} &\colhead{$z$}
  &\colhead{$\alpha$} &\colhead{$M^*$}
  &\colhead{$M_{3.6}^*$ @ $z=0.05$}
}
\startdata
\citetalias{D09}                    & 3.6$\mu$m & 0.235  & -1.12  & -25.06 $\pm$ 0.18 & -24.99 \\
\citet{babbedge}\tablenotemark{1}   & 3.6$\mu$m & 0.13   & -0.9   & -24.67 $\pm$ 0.10 & -24.72 \\
\citet{babbedge}\tablenotemark{1}   & 3.6$\mu$m & 0.38   & -1.0   & -25.07 $\pm$ 0.10 & -24.85 \\
\citetalias{lin04}\tablenotemark{2} & $K$s      & 0.043  & -0.84  & -24.02 $\pm$ 0.02 & -24.32 \\
\citetalias{lin04}\tablenotemark{2} & $K$s      & 0.043  & -1.1   & -24.34 $\pm$ 0.01 & -24.64 \\
\citet{bell}\tablenotemark{3}       & $K$s      & 0.08   & -0.77  & -24.06            & -24.32 \\
\enddata
\tablenotetext{1}{Schechter parameters were not listed in the original paper, but instead come from Table 5 of \citetalias{D09}.}
\tablenotetext{2}{\citetalias{lin04} both fit for $\alpha$ finding $\alpha = -0.84$ and fixed $\alpha = -1.1$.}
\tablenotetext{3}{No error estimate is available for \citet{bell}.}
\end{deluxetable*}

The most obvious source of systematic bias is model normalization.  As discussed above, we let the model normalization remain a free parameter in all of our fits, allowing us to quantify the error introduced by an uncertain model normalization.  We can instead use published measurements of the galaxy luminosity function at low redshift as an additional check of our results.  To allow for more direct comparisons we parameterize our fitted model normalizations by the predicted $M_{3.6}^*$ value at $z = 0.05$.  We have calculated the same value for \citetalias{lin04}, allowing us to compare directly to their results so long as we use the same value of $\alpha$.  We have also searched the literature for other measurements of the galaxy luminosity function at low redshift.  In Table \ref{tbl:normalizations} we list results from \citet{bell}, \citetalias{lin04}, \citet{babbedge}, and \citet[hereafter D09]{D09} which cover a range of values of $\alpha$, a range of redshifts, and use both $3.6\mu$m and $K$s-band.  We use our models to convert from $K$s-band to $3.6\mu$m as necessary, to correct for the predicted passive evolution from the survey redshift to $z = 0.05$, and to move the results from restframe $3.6\mu$m to observed $3.6\mu$m.  We therefore calculate observed $M_{3.6}^*$ at $z = 0.05$ for all the papers in Table \ref{tbl:normalizations}, allowing us to compare more directly to other results.

\begin{figure}
\epsscale{1.0}
\plotone{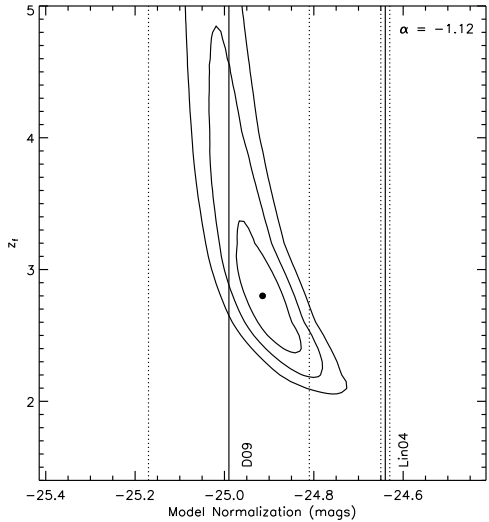}
\caption{Confidence regions for the formation epoch versus model normalization excluding the two highest redshift bins, using our results with $\alpha = -1.12$.  The solid circle denotes the best fit of $M_{3.6}^* = -24.91$, $z_f = 2.80$.  The contours represent $1, 2$ and $3\sigma$ confidence intervals for our data set.  The solid vertical and dotted lines represent the normalization and errors from \protect\citetalias{lin04} and \protect\citetalias{D09} (see Table \ref{tbl:normalizations} for the precise values).}\label{fig:zf_a_112}
\end{figure}

We begin the comparison of our fitted model normalizations to the literature by looking at the fits to our $\alpha = -1.12$ data.  While this is not our fiducial value for $\alpha$ ($-0.8$) we begin here because both \citetalias{D09} and \citetalias{lin04} have results for $\alpha \sim -1.12$, and because the differences between \citetalias{D09}, \citetalias{lin04}, and our work are instructive.  \citetalias{D09} looked at field galaxies in the ISCS.  Our survey, SDWFS, is a reimaging of the ISCS.  \citetalias{lin04} examined cluster galaxies in $K$s band.  So while the galaxies \citetalias{lin04} examines are in a similar environment to the galaxies we study, the galaxies in \citetalias{D09} were observed at the same wavelengths, with the same instrument, and some of them are even the same galaxies as ours.  Given these facts we illustrate in Figure \ref{fig:zf_a_112} the impact of the model normalization on our results.  In this Figure we show $1, 2, $ and $3\sigma$ confidence regions in $M_{3.6}^*$ versus $z_f$ space, with the best fit to the low redshift data ($z < 1.3$, $z_f = 2.80$, $M_{3.6}^* = -24.91$) marked with a filled circle.  There are two sets of solid and dashed vertical lines, representing the predicted values of $M_{3.6}^*$ at $z=0.05$ for \citetalias{lin04} and \citetalias{D09} (see Table \ref{tbl:normalizations} for the precise values).  Clearly our model normalization is in excellent agreement with predictions for \citetalias{D09}, and it is systematically brighter than \citetalias{lin04}.  The agreement with \citetalias{D09} is encouraging and, since they used the same instrument and looked at the same field, not surprising.

\begin{figure}
\epsscale{1.0}
\plotone{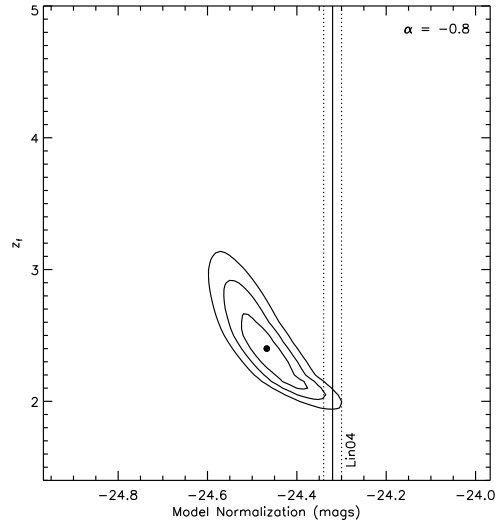}
\caption{Same as Figure \ref{fig:zf_a_112} except we have fitted our data with $\alpha = -0.8$.  The solid circle denotes the best fit of $M_{3.6}^* = -24.47$, $z_f = 2.40$.  The vertical lines represent the normalization and error from \protect\citet{lin04} with $-24.32 \pm 0.02$ and $\alpha = -0.84$.  Our fitted model normalization is consistent with the results from D09, but disagrees with L04.}\label{fig:zfconfidence}
\end{figure}

Similar to Figure \ref{fig:zf_a_112}, we show in Figure \ref{fig:zfconfidence} the $1, 2, $ and $3\sigma$ confidence regions in $M_{3.6}^*$ versus $z_f$ space for the fit to our results with $\alpha = -0.8$, for which we find $z_f = 2.40$ and $M_{3.6}^* = -24.47$.  We also show the predicted $M_{3.6}^*$ value for \citetalias{lin04} for their fit with $\alpha = -0.84$.  We note that \citet{bell} also measured $M^*$ in the local universe finding $\alpha = -0.77$.  For \citet{bell} we calculate $M_{3.6}^* = -24.32$, the same as what we find for \citetalias{lin04}.  \citet{bell} examined galaxies in the same band as \citetalias{lin04} ($K$s), so the studies with $K$s data in Table \ref{tbl:normalizations} are internally consistent.  Similar to our results for the $\alpha = -1.12$ data, the calculated $M_{3.6}^*$ value for \citetalias{lin04} is fainter than our fit, this time by $0.15$ mags.  We note that because we have fit for the model normalization our formation redshift is robust against any systematic errors in $m^*$.  Thus, even if this disagreement means that there is some unaccounted for systematic errors in $m^*$, our best fitting formation redshift will remain unchanged.  On the other hand, if we had fixed our model normalization to match \citetalias{lin04} this would have artificially forced our fit to later formation redshifts.

\begin{figure}
\epsscale{1.0}
\plotone{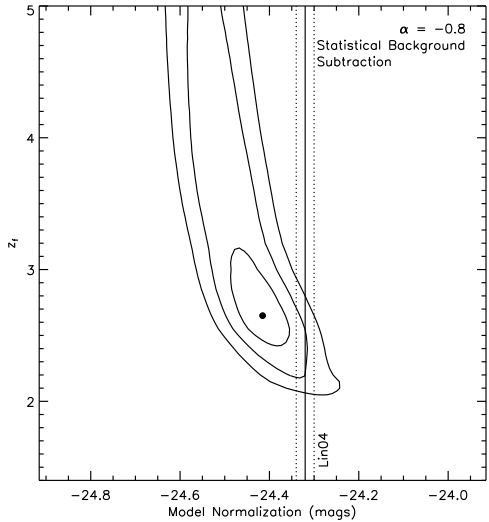}
\caption{Same as Figure \ref{fig:zfconfidence} except the models have been fit to our results using a statistical background subtraction.  The solid circle denotes the best fit of $M_{3.6}^* = -24.33$, $z_f = 2.10$.}\label{fig:zfconfidence_stat}
\end{figure}

Figure \ref{fig:zfconfidence} demonstrates that we can place tight constraints on the formation redshift even though we have left the model normalization as a free parameter.  We constrain the star formation epoch to be within $ 2.0 \lesssim z_f \lesssim 3.0$ at $3\sigma$ given our choice of model set and $\alpha$.  For comparison we include Figure \ref{fig:zfconfidence_stat} which shows the constraints on model normalization and formation redshift that result from our statistical background subtraction (Section \ref{stat_subtract}).  The best fitting formation redshift, $z_f = 2.10^{+0.06}_{-0.10}$, is consistent with the result from our photo-z selection method.  A comparison of Figures \ref{fig:zfconfidence} and \ref{fig:zfconfidence_stat} shows that while the $1 \sigma$ confidence region for the statistical background subtraction is actually smaller than for our photo-z selection method, the photo-z selection process places tighter constraints overall.  We also note that, unlike the photo-z selection method, the statistical background subtraction does in fact match the normalization of \citetalias{lin04}.  This difference is caused because the statistical background subtraction is missing the lowest redshift data point, which places the best constraints on the model normalization.  Therefore, the statistical background subtraction can have a different best fitting model normalization, despite the fact that there is good agreement between the results of our statistical background subtraction and photo-z selection method (see Figure \ref{fig:results}).

\subsubsection{Faint End Slope ($\alpha$)\label{alpha_errors}}

Since we have assumed a value of $\alpha$ and fixed it in our fits, it is important to examine the impact of $\alpha$ on our results.  This is particularly important because there is a strong correlation between $m^*$ and $\alpha$.  We have fit our data using a variety of values for $\alpha$, $-0.6$, $-0.8$, $-1.0$, and $-1.12$.  This covers the full range of values for $\alpha$ listed in Table \ref{tbl:normalizations} and commonly found in the literature.  As can be seen in Table \ref{tbl:results} the behavior of $m^*$ as a function of $\alpha$ is well behaved and smooth.  In general a change in $\alpha$ of 0.1 corresponds to a systematic change in $m^*$ of $\sim 0.1$ mags.  The change in $m^*$ is largely constant as a function of redshift, and so the observed shape of the evolution in the luminosity function changes only slightly as a function of $\alpha$.

We fit models to our results for $-0.6 < \alpha < -1.12$, and list the results in Table \ref{tbl:systematics}.  The most important fact is that the best fitting formation redshift is fairly robust against changes of $\alpha$.  Changing $\alpha$ has only a small impact on the observed evolution of $m^*$, and as a result changing $\alpha$ primarily changes the fitted model normalization.  The best fitting formation redshift varies from $z_f = 2.80$ for $\alpha = -1.12$ to $z_f = 2.40$ for $\alpha = -0.6$ while at the same time our fitted $m^*$ values systematically shift by $\sim 0.6$ mags over this same range of $\alpha$.

We already mentioned in Section \ref{model_errors} that our fitted model normalization for $\alpha = -1.12$ is in good agreement with \citetalias{D09}, but disagrees with \citetalias{lin04}.  We note now that we also have good agreement with \citet{babbedge}.  For their results we calculate $M_{3.6}^* = -24.72 \pm 0.1$ and $M_{3.6}^* = -24.85 \pm 0.1$ for their $z = 0.13$ and $z = 0.38$ data points, which have $\alpha = -0.9$ and $\alpha = -1.0$.  For $\alpha = -1.0$ we find $M_{3.6}^* = -24.75 \pm 0.04$, which is within the errors for both.  This means that our fitted model normalizations agree with both $3.6\mu$m surveys (\citetalias{D09} and \citealt{babbedge}) but disagree with both $K$s surveys (\citetalias{lin04} and \citealt{bell}) over a range of values of $\alpha$.

We use the fitted formation redshifts across the range of $\alpha$ studied in this paper as an estimate for the systematic error introduced by $\alpha$.  Therefore we find that uncertainties in $\alpha$ introduce an uncertainty into our best fitting formation redshift of $\Delta z_f = ^{+0.40}_{-0.05}$.

Finally we note that the necessity of fixing $\alpha$ in all redshift bins introduces an important limitation -- we clearly cannot measure any evolution in $\alpha$.  Moreover, the coupling between $m^*$ and $\alpha$ means that if there is any evolution in $\alpha$ then our fitted values for $m^*$ will not agree with what would be measured using the proper value of $\alpha$.  However this has only a small impact on our ability to measure the formation redshift for massive galaxies.  As long as we choose a consistent value for $\alpha$ at all redshifts we can accurately measure any relative change in the bright end of the LF as a function of redshift, and therefore we can still properly measure the formation redshift of massive galaxies.

\subsubsection{Stellar Population Synthesis Modeling (SPS)\label{sps_errors}}

Our choice of SPS model set is clearly an important ingredient in our analysis.  To assess the sensitivity of our results to this ingredient we redo our $\chi^2$ minimization using the models of \citetalias{BC03} and \citetalias{CB07}.  For \citetalias{BC03} we find $z_f = 1.95^{+0.21}_{-0.07}$, $M^*_{3.6} = -24.57^{+0.04}_{-0.05}$ and for \citetalias{CB07} we find $z_f = 2.70^{+0.19}_{-0.19}$, $M^*_{3.6} = -24.57 \pm 0.04$ (compared to our fiducial fit of $z_f = 2.40^{+0.16}_{-0.18}$, $M^*_{3.6} = -24.47 \pm 0.04$).  These results are also summarized in Table \ref{tbl:models} where we list both best fitting formation redshift, model normalization, and $\chi_{\nu}^2$ for all three model sets.  We use the differences between these results to estimate the systematic error introduced by our choice of model ($\Delta z_f = ^{+0.30}_{-0.45}$).

We also investigate the effect of common sources of systematic biases in the models, such as the parameterization of the IMF and metallicity.  If we use a Salpeter IMF instead of Kroupa the best fitting formation redshift moves to $z_f = 2.15$.  Using both a super-solar and sub-solar metallicity moves our results to earlier formation redshifts ($z_f = 2.8$ and $z_f = 3.00$, respectively).  These results are also summarized in Table \ref{tbl:models}.

\begin{figure}
\epsscale{1.0}
\plotone{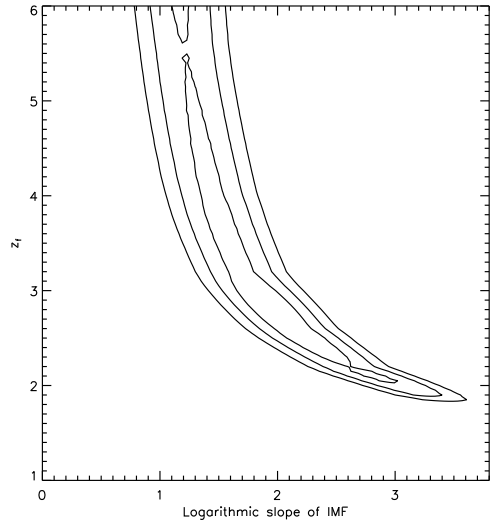}
\caption{1, 2, and 3$\sigma$ confidence regions for the formation epoch versus the logarithmic slope of the IMF, excluding the two highest redshift bins.  For these fits we have used the data with $\alpha$ fixed to $-0.8$ and have fixed the model normalization to $M_{3.6}^* = -24.47$, taken from our fiducial fit to the data.  In these units a Salpeter slope is 2.35.}\label{fig:fit_imf}
\end{figure}

Recently \citet{conroy} performed a thorough investigation of potential systematic uncertainties in SPS modeling.  For our purposes the most important uncertainty they found was the logarithmic slope of the IMF around $1M_\odot$.  For old stellar populations the mass of the main sequence turn off (MSTO) is about $1M_\odot$, and the slope of the IMF at the MSTO is an important factor for determining the evolution of the luminosity of a stellar population.  \citet{conroy} estimates the impact that the uncertainty in slope of the IMF near $1M_\odot$ has on SPS modeling and concludes that the level of systematic bias in the predicted evolution of the luminosity function in the current generation of stellar models could be as large as $\sim0.4$ mags per unit redshift.  To account for this we add to our models a simple prescription (Conroy, private communication) to describe the relationship between the slope of the IMF at $1M_\odot$ and the evolution of the models (dmag/d$z$) in $3.6$ and $4.5\mu$m, and fit for $z_f$ as a function of the IMF slope.  For the fit we use our data with $\alpha$ fixed at $-0.8$ and fix the model normalization to $M_{3.6}^* = -24.47$.  The result is shown in Figure \ref{fig:fit_imf} which plots $1, 2,$ and $3\sigma$ confidence intervals around the best fitting combination of IMF slope and $z_f$.  A Salpeter slope in these units is 2.35.  Clearly there is a strong degeneracy between the slope of the IMF and the best fitting formation redshift.  If the slope of the IMF near $1M_\odot$ is flatter at high redshift, then the actual formation redshift can be substantially earlier.

In this context it is important to consider the work of \citet{vanDokkum08} who examined the evolution of both the color and mass-to-light ratio of cluster galaxies.  Because the luminosity evolution of galaxies is strongly dependent upon the IMF and the color evolution is not, \citet{vanDokkum08} was able to place joint constraints on the IMF and the formation redshift of the cluster galaxies in his sample.  Looking at just the evolution of the mass-to-light ratio and assuming a Salpeter IMF he found good agreement to models with $z_f = 2.0$, similar to what we have found from the luminosity evolution of our sample.  However this same model could not fit the color evolution in his sample for which an earlier formation redshift, $z_f = 6.0$, was necessary.  \citet{vanDokkum08} found that the color and mass-to-light evolution could only be simultaneously fit by a flat IMF and an early formation redshift ($z_f = 6.0$).  They conclude that the slope of the IMF at high redshift is much flatter than a Salpeter IMF, which if true means that our fitted formation redshift ($z_f = 2.4$) is really only a lower limit on the formation redshift, as per Figure \ref{fig:fit_imf}.  As noted in \citet{conroy}, this degeneracy between the IMF and the formation redshift is common to all studies that rely on the luminosity evolution of a stellar population.  Clearly then the slope of the IMF at high redshift is the most important systematic uncertainty in this study, as well as similar ones in the past.  With a sufficiently flat IMF, our results can fit arbitrarily early formation redshifts.

The other major uncertainties discussed in this paper introduce relatively small systematic errors: $\Delta z_f = ^{+0.16}_{-0.18}$ (model normalization), $\Delta z_f = ^{+0.40}_{-0.05}$ ($\alpha$), and $\Delta z_f = ^{+0.30}_{-0.45}$ (model set).  Added in quadrature this gives a total error of $\Delta z_f = \pm 0.5$.  While our error bars leave room to fit earlier formation redshifts, it is clear that a flat IMF can lead to even earlier formation redshifts.  This could account for any apparent disagreement between our results and the work of \citet{eisenhardt} who looked at the color evolution of this same sample of clusters.  \citet{eisenhardt} find $z_f = 3$ with a trend towards earlier formation redshifts for higher redshift clusters ($z > 1$).  If the IMF is flatter at higher redshifts, then the color evolution of the clusters should point to an earlier formation redshift than the luminosity evolution.  As with \citet{vanDokkum08} we can use this fact to place joint constraints on the formation redshift and the IMF by combining both the luminosity and color evolution of our clusters.  This will be an important goal for our future work.

\subsubsection{Fitting Errors\label{fitting_errors}}

\begin{figure}
\epsscale{1.0}
\plotone{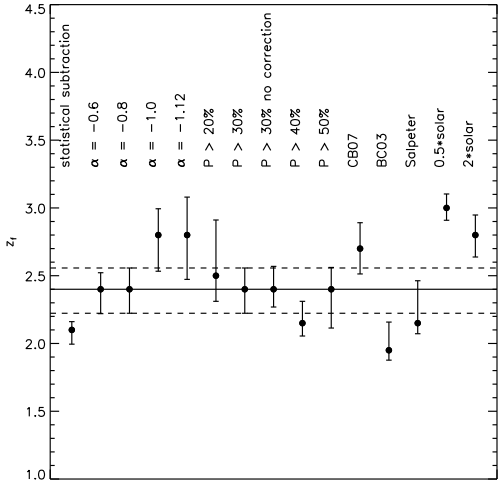}
\caption{Best fitting formation redshift for fitting systematics and model systematics as given in Tables \ref{tbl:systematics} and \ref{tbl:models}.  The solid and dashed lines represent our chosen fiducial fit and errors of $2.40^{+0.16}_{-0.18}$.  We note that both the $\alpha = -0.8$ data point and the P $>$ 30\% data point represent the same fit to the same data (our fiducial fit).  It has been included twice to illustrate the dependence of our results on both $\alpha$ and the cut probability.}\label{fig:systematics}
\end{figure}

\begin{figure}
\epsscale{1.0}
\plotone{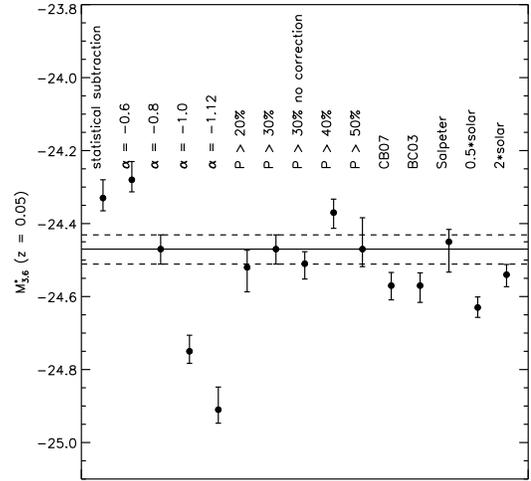}
\caption{The same as Figure \ref{fig:systematics} except for the best fit model normalizations ($M_{3.6}^*$) and errors for the various model systematics.  The solid and dashed lines represent our chosen fiducial fit with $-24.47 \pm 0.04$.  We note that both the $\alpha = -0.8$ data point and the P $>$ 30\% data point represent the same fit to the same data (our fiducial fit).  It has been included twice to illustrate the dependence of our results on both $\alpha$ and the cut probability.}\label{fig:systematics_n}
\end{figure}

Finally we study in detail the impact of the systematic biases introduced into our fitting process which were discussed in Section \ref{lf_errors}.  We apply the same $\chi^2$ minimization process to each variation of our fitting procedure using the \citetalias{M05} models with $\alpha$ fixed at $-0.8$ and calculate the best fitting formation redshifts, model normalizations, and errors.  We list the results in Table \ref{tbl:systematics} along with $\chi^2_\nu$ for each fit.  Figures \ref{fig:systematics} and \ref{fig:systematics_n} directly compare all our systematic biases graphically by showing the results from Tables \ref{tbl:systematics} and \ref{tbl:models}.  We plot the best fitting formation redshift and errors for both fitting systematics and model systematics in Figure \ref{fig:systematics}.  The best fitting normalizations and errors are shown in Figure \ref{fig:systematics_n}.  Considering the variety of systematic uncertainties that we have investigated, the general agreement found in Figure \ref{fig:systematics} was not a foregone conclusion.  All of our best fits fall in the range of $1.9 < z_f < 3.0$.

Could the good agreement between the various systematic biases be an artifact of our choice to let the normalization float?  Any signs of a discrepancy might show up in the model normalization instead of the formation redshift.  However Figure \ref{fig:systematics_n}, which compares the fitted model normalizations in the same way as Figure \ref{fig:systematics}, demonstrates that this is not the case.  The largest outliers correspond to changes in $\alpha$.  As discussed in Section \ref{alpha_errors} we expect large changes in the normalization for different values of $\alpha$ because of the strong correlation between $\alpha$ and $m^*$.  Our consistency check with results from the literature (see Sections \ref{norm_errors} and \ref{alpha_errors}) shows that these normalizations are consistent with their expected values from other $3.6\mu$m studies, and so there is no evidence of a problem.

Overall the consistency found in Figures \ref{fig:systematics} and \ref{fig:systematics_n} is quite remarkable, and is a testimony to the robustness of our results and the quality of our data.  At this point the most important systematic uncertainties are the slope of the IMF near 1 M$_\odot$ and our assumed value of $\alpha$.  Future work which desires to build upon our results should concentrate on constraining $\alpha$ and the IMF at high redshift.

\begin{deluxetable*}{cccc}
\tablecaption{Impact of Fitting Systematics.\label{tbl:systematics}}
\tablewidth{0pt}
\tablehead{
  \colhead{Systematic Uncertainty} & \colhead{$z_f$} & \colhead{$M_{3.6}^*$} & \colhead{$\chi_{\nu}^2$}
}
\startdata
Statistical Background Subtraction     & $2.10^{+0.06}_{-0.10}$ & $-24.33^{+0.05}_{-0.04}$ & 0.91 \\
Fix $\alpha = -0.6$                    & $2.40^{+0.12}_{-0.18}$ & $-24.28^{+0.05}_{-0.03}$ & 1.33 \\
Fix $\alpha = -0.8$                    & $2.40^{+0.16}_{-0.18}$ & $-24.47^{+0.04}_{-0.04}$ & 1.50 \\
Fix $\alpha = -1.0$                    & $2.80^{+0.19}_{-0.27}$ & $-24.75^{+0.04}_{-0.03}$ & 0.94 \\
Fix $\alpha = -1.12$                   & $2.80^{+0.28}_{-0.33}$ & $-24.91^{+0.06}_{-0.04}$ & 0.79 \\
Cut galaxies with probability $< 20\%$ & $2.50^{+0.41}_{-0.19}$ & $-24.52^{+0.05}_{-0.07}$ & 1.17 \\
Cut galaxies with probability $< 30\%$ & $2.40^{+0.16}_{-0.18}$ & $-24.47^{+0.04}_{-0.04}$ & 1.50 \\
Cut $< 30\%$ with no LF correction     & $2.40^{+0.17}_{-0.13}$ & $-24.51^{+0.03}_{-0.04}$ & 1.56 \\
Cut galaxies with probability $< 40\%$ & $2.15^{+0.16}_{-0.10}$ & $-24.37^{+0.04}_{-0.04}$ & 1.32 \\
Cut galaxies with probability $< 50\%$ & $2.40^{+0.16}_{-0.29}$ & $-24.47^{+0.09}_{-0.05}$ & 2.72 \\
\enddata
\end{deluxetable*}

\begin{deluxetable*}{ccccccc}
\tablecaption{Impact of Model Choice.\label{tbl:models}}
\tablewidth{0pt}
\tablehead{
  \colhead{Model Set} & \colhead{IMF} & \colhead{Metallicity} & \colhead{$z_f$} & \colhead{$M_{3.6}^*$} & \colhead{$\chi_{\nu}^2$}
}
\startdata
\citetalias{M05}  & Kroupa   & Solar     & $2.40^{+0.16}_{-0.18}$ & $-24.47^{+0.04}_{-0.04}$ & 1.50 \\
\citetalias{M05}  & Salpeter & Solar     & $2.15^{+0.31}_{-0.08}$ & $-24.45^{+0.03}_{-0.08}$ & 1.57 \\
\citetalias{M05}  & Salpeter & 0.5*Solar & $3.00^{+0.10}_{-0.09}$ & $-24.63^{+0.03}_{-0.03}$ & 1.75 \\
\citetalias{M05}  & Salpeter & 2*Solar   & $2.80^{+0.15}_{-0.16}$ & $-24.54^{+0.03}_{-0.03}$ & 1.66 \\
\citetalias{CB07} & Chabrier & Solar     & $2.70^{+0.19}_{-0.19}$ & $-24.57^{+0.04}_{-0.04}$ & 0.95 \\
\citetalias{BC03} & Chabrier & Solar     & $1.95^{+0.21}_{-0.07}$ & $-24.57^{+0.04}_{-0.05}$ & 0.60 \\
\enddata
\end{deluxetable*}

\section{Mass Assembly}

\subsection{Low Redshift}

\begin{figure}
\epsscale{1.0}
\plotone{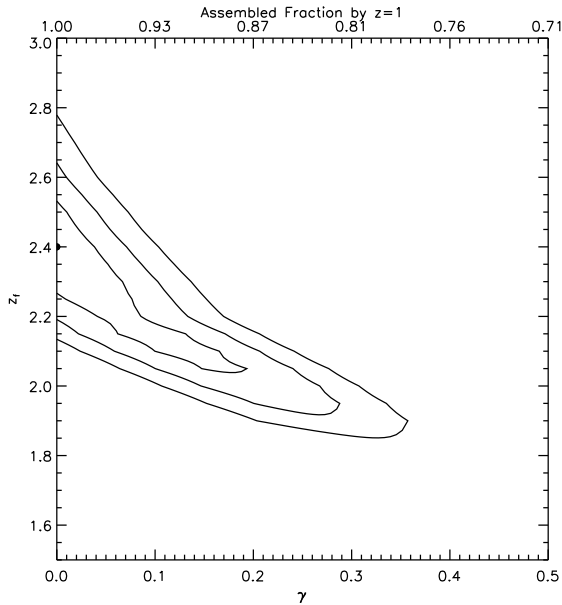}
\caption{1, 2, and 3$\sigma$ confidence regions for the formation epoch versus $\gamma$, which parameterizes the rate of mass assembly in low redshift galaxies.  In this model galaxies are assumed to lose mass at the rate of $(1+z)^\gamma$ out to higher redshifts.  The bottom axis shows the values of $\gamma$ fitted, and the top axis shows the fraction of the final galaxy mass that is assembled at $z = 1.0$ for a given value of $\gamma$.  The solid circle shows our best fit of $z_f = 2.40$, $\gamma = 0$.}\label{fig:fit_assembly}
\end{figure}

We also examine the impact of ongoing mass assembly in our low redshift bins.  While our results are well fit by models for passively evolving stellar populations, this does not rule out the possibility of galaxy assembly.  To test for the presence of mass assembly in our low redshift galaxies, we create a simple model for luminosity evolution in which the same models for passive evolution are used but an additional dimming factor of $(1+z)^\gamma$ is introduced to account for the breakup of galaxies at higher redshift.  We fit to the $z < 1.3$ redshift bins, use the data with $\alpha = -0.8$, and fix the model normalization to $M_{3.6}^* = -24.47$.  We find the best fitting combination of formation redshift and $\gamma$ ($z_f = 2.40, \gamma = 0$) and show the 1, 2, and 3$\sigma$ contours for this fit in Figure \ref{fig:fit_assembly}.  Our best fit demonstrates clearly that we have no need for additional mass assembly (at least for this toy model), in agreement with past work \citep{strazzullo,depropris}.  However it is also clear that our results can fit a small amount of mass assembly.  $\gamma \sim 0.2$ is consistent at $1\sigma$, and $\gamma \sim 0.35$ is consistent at $3\sigma$.  These correspond to $\sim 90\%$ and $\sim 75\%$ of the final galaxy mass being assembled by $z = 1.0$.  Considering that we have constructed only a very simple toy model for mass assembly at low redshifts, these numbers should not be considered authoritative.  However this should make it clear that there is room for growth of massive galaxies at low redshift, albeit not much.

Finally we note that the presence of mass assembly at low redshift has an impact on the best fit formation redshift.  Since mass assembly causes the galaxies to grow fainter out to higher redshift, accounting for mass assembly causes the models to grow steeper and therefore fits a later formation redshift.  As can be seen in Figure \ref{fig:fit_assembly} our data and toy model could allow for formation redshifts as late as $z_f \sim 2.1$ at $1\sigma$, or $z_f \sim 1.9$ at $3\sigma$.

\subsection{High Redshift\label{mass_assembly}}

We noted in Section \ref{model_comparison} that our best fit model deviates (in a statistically significant way) from the two highest redshift data points.  There are two possible explanations for this phenomenon: either a systematic error is impacting the high redshift data or this is a signature of galaxy assembly.  We have already discussed a number of possible systematic errors in both the fitting process (Section \ref{lf_errors}) and in our models (Section \ref{model_errors}).  We find that the disagreement between the models and data at high redshift is a consistent feature found for all the variations of model and fitting parameters listed in Tables \ref{tbl:systematics} and \ref{tbl:models}.  Moreover we find that this disagreement remains for all of the variations of the IMF slope discussed in Section \ref{sps_errors}, and is robust against changes in redshift bin sizes and locations.  Evolution in $\alpha$ is also unlikely to explain this result, as it would require the LF to be significantly steeper at high redshift.

A more subtle systematic uncertainty could be the cause of this high redshift deviation.  For instance, our photometric redshifts begin to degrade at $z \sim 1.5$, where the models and data begin to disagree.  Also, the most distant spectroscopically confirmed cluster is at z=1.48, which is near the point where the models no longer match the data.  It is possible that the false positive rate for our cluster search grows rapidly at high redshift, at too high of a redshift for the problem to be detected with our data.  However it is not clear how either of these two possibilities would cause the data to grow systematically fainter by $\sim$1 magnitude.  We have performed a simple test using our highest redshift data bin that still matches the models ($z = 1.24$).  We scattered all the photo-z probability distributions to simulate what would happen if the photo-z errors were underestimated by a factor of $\sim3$, and we have moved galaxies from random redshifts into the bin to simulate the presence of catastrophic failures in the photometric redshifts.  We added enough such that at least $5\%$ of the galaxies in the bin are catastrophic failures.  Despite these drastic changes the fitted value of $m^*$ for the $z = 1.24$ bin only scattered within its error bars.  This suggests that random errors in the photometric redshifts cannot be responsible for such a large systematic offset.

Systematic errors in the photometric redshifts are more likely to cause this discrepancy, but do not seem adequate to account for it.  If the photometric redshifts are systematically underfitting the redshift of high-z galaxies, then the galaxies will appear fainter than expected because the selected galaxy population will be at a higher redshift than claimed.  This would also explain why there is no apparent discrepancy found in the colors of the high redshift data (see Figure \ref{fig:resultscolor}), since there is little color evolution in the models for $z \gtrsim 1.5$.  However in order to explain the $\sim1$ mag offset, the galaxies in the $z = 1.5$ bin would have to have a mean redshift of $z \gtrsim 2.0$.  Such a large error seems unlikely, especially since there is no evidence of $z = 1.5$ galaxies ending up in the $z = 1.24$ redshift bin.  Similarly, we see no reason why an increasing false positive rate would cause a systematic shift to fainter magnitudes for our data.  Finally we note that in Figure \ref{fig:results} the results of our statistical background subtraction also show evidence for this same disagreement at high redshift.  This is particularly significant because photometric redshifts do not factor into the statistical background subtraction, and because it should in general be less prone to systematic errors than our photo-z selection method.  We conclude that there is no strong evidence that systematic errors cause the disagreement between data and models at high redshift.  Further work is required to investigate the cluster LF in detail at high redshift, and for now we will discuss the implications of our result if these deviations are real.

The discrepancy at high redshift is easiest to explain by ongoing mass assembly.  One point in favor of growth through mergers is the lack of color evolution, relative to the models, for the high redshift clusters.  As can be seen in Figure \ref{fig:resultscolor}, the color of an $m^*$ galaxy matches the models over the entire redshift range of our sample, despite a $\sim1$ mag deviation in apparent magnitude at high redshift.  As long as the galaxy assembly process doesn't appreciably change the luminosity weighted mean age of the stellar populations the color of the clusters will continue to match the models, even while the luminosity increases due to the increase of mass.

If we assume that all of the deviation at high redshifts comes from galaxy assembly and that there is no significant assembly in the low redshift bins, then we can estimate the fraction of the final galaxy mass that is assembled in our last two bins.  We simply assume that the fractional decrease in luminosity relative to the models correlates directly with the fraction of mass ``lost" to assembly.  This implies that the galaxies grow by a factor of $\sim 2-4$ from $1.3 < z < 2.0$.

Ours is the first survey to track the evolution of the luminosity function of cluster galaxies to such high redshifts, so there is little in the literature to which we can directly compare our result for clusters. This result does however appear to be at odds with what is found in the field. \citet{ilbert10} for example examine the stellar mass function of quiescent field galaxies over this same redshift range, finding little evolution in $M^*$ (the characteristic mass of field galaxies) at high redshift, $1.0 < z < 2.0$. A number of other studies have also found that the high mass end of the field population is fully formed by $z \sim 2$ (e.g. \citealt{perez08} and \citealt{marchesini09}). It would be surprising for the cluster galaxies to still be assembling when the massive field galaxies are in place; however we are unable to identify any systematic error that would qualitatively alter our result.  Our z < 1.3 results indicate that the $z\sim 2$ era was likely an active one in cluster galaxy formation.  Further investigation of the properties of galaxies in rich environments at $z \sim 2$ is needed.

\section{Summary and Conclusions}

We measure the evolution of the $3.6$ and $4.5\mu$m luminosity functions for massive cluster galaxies using results from the SDWFS.  We find that for $z \lesssim 1.3$ our data are consistent with passively evolving stellar populations with a mean formation redshift of $z = 2.40^{+0.16}_{-0.18}$ assuming a Kroupa IMF and $\alpha = -0.8$.  We investigate a number of possible systematic errors that could be biasing our best fitting formation redshift, and summarize the most important here:

1) We can only measure $\alpha$ in the lowest redshift bin of our data, and so must fix $\alpha$ in our fits.  Our choice of $\alpha$ introduces an uncertainty in our formation redshift of $\Delta z_f = ^{+0.40}_{-0.05}$.  Moreover fixing $\alpha$ means that we are not sensitive to any evolution in $\alpha$.  However our final formation redshift should not have a large dependence on any evolution in $\alpha$ because our ability to measure the location of the bright end of the LF is independent of what the faint end might be doing.  As long as we choose a consistent value for $\alpha$ at all redshifts we can accurately measure any relative change in the bright end of the LF as a function of redshift, and therefore we can still properly measure the formation redshift of massive galaxies.

2) Our choice of model set introduces a systematic uncertainty of $\Delta z_f = ^{+0.30}_{-0.45}$ into our results if we consider \citetalias{CB07} and \citetalias{BC03} relative to \citetalias{M05}.

3) The uncertainty in the slope of the IMF near $1M_\odot$ introduces a bias into the evolution of the luminosity function on the order of $\sim0.4$ mags per unit redshift.  This bias is large enough that a sufficiently flat IMF can result in arbitrarily early formation redshifts.  Given that \citet{vanDokkum08} finds evidence for a flat IMF based on joint constraints from the color and mass-to-light evolution of their cluster galaxies, a flatter IMF is certainly plausible.  This degeneracy between the formation redshift and IMF is best broken by a combination of color and luminosity evolution (as was done by \citealt{vanDokkum08}), a possibility which we intend to investigate with this cluster sample in future work.

We generate a simple toy model to explore the growth of massive galaxies at low redshift, and find that our data are best fit with no additional growth through assembly for $z < 1.3$.  However even for the simple model we generate, our data are still consistent with some growth at low redshift.  We conclude that even though our galaxies are well matched by passively evolving stellar populations at low redshift, this does not mean that there is no mass assembly of low redshift galaxies.

Our best fit model disagrees significantly ($\sim 5\sigma$) with the highest redshift data ($z = 1.3 - 2.0$).  We cannot completely rule out the possibility that this deviation is caused by systematic uncertainties such as errors in the photometric redshifts at high redshift or a high false detection rate of clusters.  However we explore the implications of what this deviation would mean if it is real, and conclude that it could be a signature of ongoing mass assembly.

\acknowledgments

We would like to thank Gustavo Bruzual and Stephane Charlot for sharing their latest SSP model set.  We would also like to thank Charlie Conroy for a helpful discussion and for generating models specifically for this work.  Finally we are grateful to the anonymous referee for suggestions and input that have greatly improved this work.

This paper is based upon work supported by the National Science Foundation under grant AST-0708490.  Part of this work was performed under the auspices of the U.S. Department of Energy by Lawrence Livermore National Laboratory in part under Contract W-7405-Eng-48 and in part under Contract DE-AC52-07NA27344.  The work of PE and DS was carried out at Jet Propulsion Laboratory, California Institute of Technology, under a contract with NASA.  This work is based on observations made with the Spitzer Space Telescope, which is operated by the Jet Propulsion Laboratory, California Institute of Technology under a contract with NASA. Support for this work was provided by NASA through an award issued by JPL/Caltech.

\bibliographystyle{apj}
\bibliography{galaxy_lf}

\end{document}